# A Parametric Rate-Distortion Model for Video Transcoding

## Maedeh Jamali, Nader Karimi, Shadrokh Samavi, Shahram Shirani

**Abstract** Over the past two decades, the surge in video streaming applications has been fueled by the increasing accessibility of the internet and the growing demand for network video. As users with varying internet speeds and devices seek high-quality video, transcoding becomes essential for service providers. In this paper, we introduce a parametric rate-distortion (R-D) transcoding model. Our model excels at predicting transcoding distortion at various rates without the need for encoding the video. This model serves as a versatile tool that can be used to achieve visual quality improvement (in terms of PSNR) via trans-sizing. Moreover, we use our model to identify visually lossless and near-zero-slope bitrate ranges for an ingest video. Having this information allows us to adjust the transcoding target bitrate while introducing visually negligible quality degradations. By utilizing our model in this manner, quality improvements up to 2 dB and bitrate savings of up to 46% of the original target bitrate are possible. Experimental results demonstrate the efficacy of our model in video transcoding rate distortion prediction.

## 1. Introduction

The growth of video streaming applications, encompassing various domains such as online conferences, group gaming, and virtual worlds has increased the demand for video streaming. A substantial increase in Internet Protocol Television (IPTV) services has been witnessed during recent years. According to Cisco's reports [1], network video traffic will increase over the next five years and a huge number of video contents will be generated per second. The success of these video services will be directly related to their user satisfaction. Achieving seamless live streaming depends on critical factors, such as ample network bandwidth, predictive video quality optimization for network planning and service optimization [2-4].

Video consumption has undergone a significant transformation in recent years, with a shift from traditional TV systems to streaming on a variety of devices, including smartphones, laptops, and desktops. The Global Internet Phenomena Report [5] highlights video streaming as one of the most prevalent Internet applications, dominating 80% of global Internet traffic [6].

With the increase in the demand for online video applications, the challenge of ensuring consistent Quality of Experience (QoE) for users with limited network capacities becomes more pronounced. As more users turn to mobile networks the need for adaptable streaming qualities becomes increasingly evident. Adaptive BitRate (ABR) streaming has emerged as a solution to address these challenges. ABR enables real-time bitrate adaptation during video sessions by transcoding the source video into multiple representations at various bitrates. It allows a rate adaptation during video session by transcoding source video into several representations in various bitrates [7-8]. This adaptation involves converting the original video into a compatible resolution, frame rate, video codec, and network bandwidth to align with viewer display devices [9]. Transcoding encompasses various processes, including:

- Transcoding: At a high level, this involves receiving compressed content, decompressing it, making alterations, and recompressing it.
- Trans-rating: This pertains to changing the bitrate of the video or audio and converting it into one or more lower bitrates to enable ABR streaming.
- Trans-sizing: This involves resizing the video frame, such as reducing it from a resolution of 1080p to 720p or 480p.

In video streaming, transcoding is necessary when there are resource disparities between the sender and receiver [10].

The International Telecommunication Union's (ITU) Study Group 12, under the Telecommunication Standardization Section, has standardized ITU-T Recommendation G.1071 [11] to address video-streaming services in planning models. Additionally, ITU-T Recommendations P.1203 [12] establish a parametric video quality estimation module for assessing visual quality in adaptive streaming video services and progressive downloads using the H.264 video codec. ITU-T SG12 is presently engaged in the AVHD-AS/P.NATS phase 2 project, aiming to develop a more comprehensive parametric model capable of meeting the requirements of modern video streaming applications [13].

For large video streaming platforms like YouTube, optimizing streaming costs while delivering the highest video quality with minimal bitrates, is crucial. This optimization directly impacts the platform's operating costs, data expenses, and the QoE of end-users. Different types of video content, ranging from sports and TV shows to animations and gaming [14], are hosted on these platforms. Each content type exhibits unique R-D characteristics due to varying content complexity. A parametric model capable of capturing the R-D behavior of video segments, regardless of content, holds great potential for efficiently addressing these challenges:

- Efficiency: By predicting R-D categories and sharing encoding instructions among contents with similar R-D characteristics, computational efficiency is improved.
- Resolution and Bitrate Selection: The model can propose optimal resolutions and bitrates tailored to user needs, enhancing overall quality.

In this paper, we propose a parametric model for transcoding. As we can see in Fig. 1, our model can be used to make RD-based decisions about the target resolution (trans-sizing) and target bit rate (trans-rating) for an ingest video during transcoding.

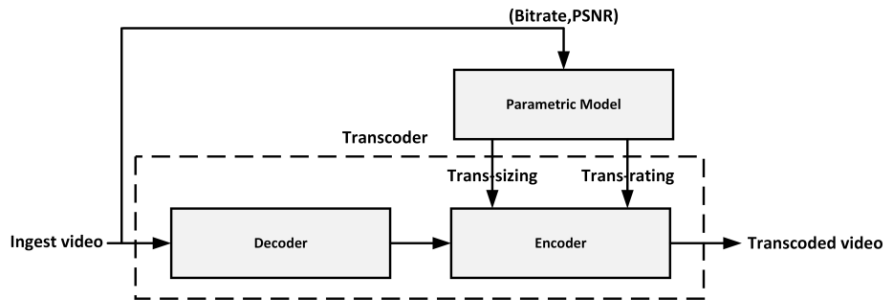

**Fig. 1** Block diagram of the transcoding process in connection with our parametric model

During model formation we cluster the R-D curve of videos from different types and resolutions into distinct groups. The number of clusters required to model the dissimilarity in the R-D characteristics across the videos is much less than the size of the dataset. Since all videos in a cluster have similar R-D characteristics, we can make RD decisions based on clusters rather than individual videos. The behavior of each cluster is represented by its centroid. By assigning an ingest video to a cluster, our model provides the R-D characteristics of the video at different rates and resolutions. This is an invaluable source of information for transcoder to make decision about the target resolution and bit rate.

The significant contributions of our proposed method are as follows:

- Introduction of a parametric model for R-D behavior in video transcoding to predict suitable bitrates and resolutions for ingest videos.
- Utilization of a diverse dataset of video types for model formation, decoupled from specific content types.
- Our parametric model can be used by a transcoder for bitrate and resolution assignment /optimization.
- Elimination of the computationally intensive and time-consuming processes associated with conventional transcoding optimization which involves coding a video at various bitrates
- Our model can be used to make decision regarding the target resolution (trans-sizing) and bitrate (trans-rating) which can introduce significant quality improvements and/or bitrate savings.

The remainder of this paper is organized as follows: Section 2 presents a comprehensive literature review. Section 3 delves into the details of the proposed method. Section 4 discusses the experimental results, while Section 5 provides the paper's conclusion.

## 2. Literature Review

The upward trend of digital visual media is continuing. An incremental number of applications are being proposed for Internet video streaming, and video entertainment that includes video content providers such as YouTube, Netflix, Twitch, and Amazon Video, as well as video-centric mobile applications including Snapchat, Google Hangouts, and Instagram. There is a growing demand for high-quality video content on diverse devices with different bandwidth capacities and display resolutions [4].

There are multiple stages before and during the delivery of streamed digital video content to the viewership [15].

Network capacity limitations, network fluctuations, and the limits of bandwidth can lead to unstable network conditions, that on the client side can interrupt video playback by rebuffering or stalling events. Such kind of problems will impact the satisfaction of end users and lead to user attrition and viewer desertion [16].

A video stream is composed of several distinct sequences. Each of these sequences is further subdivided into multiple groups of pictures (GOP). At the start of each GOP, there is sequence header information. A GOP, in essence, represents a series of frames that are related to the same scene or content within the video [17]. Since each GOP within a video stream can be handled as a separate and self-contained unit, the transcoding operation is often performed at the GOP level [18]. To deliver a superior video streaming experience, whether through Video on Demand (VOD) platforms such as YouTube and Netflix or live-streaming services like Livestream, it is essential to adapt the video content to align with the specifications of the viewers' playback devices. This adaptation necessitates the conversion of the original video to compatible settings, including resolution [19][20], frame rate [21], video codec [22][23], toolsets such as Bi Predictive pictures, Quarter Pixel motion vectors [24], [25]and network bandwidth, in order to ensure optimal playback on the viewers' display devices [6]. These conversions are carried out on all GOPs of a video and are commonly referred to as video transcoding [15], a computationally intensive and time-consuming operation that involves encoding and decoding video data to match the specific requirements of the target devices [16]. The process of transcoding involves two main steps: first, decoding the GOPs, and then it re-encoding them into the desired format. Consequently, the total time required for transcoding is the combined time it takes for both decoding and re-encoding [26]. Figure 2 illustrates the classification of various video transcoding methods. Homogeneous transcoding entails adjusting specific video features (such as resolution, bit rate, or frame rate) while remaining within the same video compression format. It ensures seamless compatibility across different devices and platforms. In contrast, heterogeneous transcoding involves a more significant transformation—it changes the entire video compression format itself. This adaptation is necessary when the existing format isn't supported by the chosen video container or when the target device requires a different format [27].

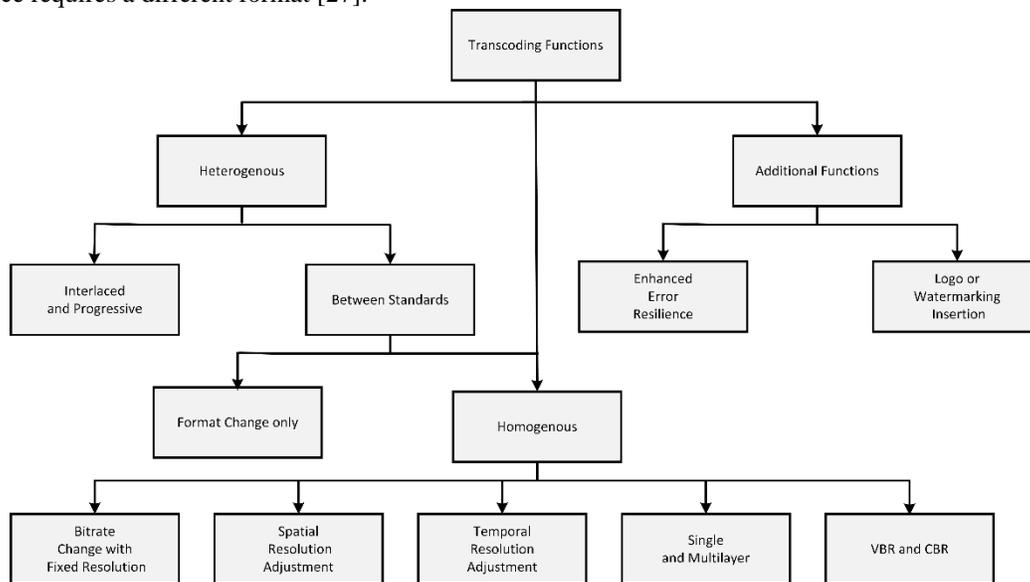

**Fig. 2** Types of video transcoding

In the following, more details about different transcoding operations are provided.

- Bitrate adjustment (Trans-rating): A process involving bit rate transcoding within the same video format, plays a crucial role in digital video adaptation and distribution. It serves two main purposes, (1) network bandwidth adaptation and (2) device-specific adaptation [28]. In the first case, when transmitting video over networks with varying bandwidth, temporary capacity issues can arise. If the video's bitrate exceeds the available network bandwidth, it leads to visual artifacts due to packet loss. To mitigate these distortions, the video stream must be scaled down to a lower rate in a controlled manner. Trans-rating ensures that bitrate adaptation aligns optimally with network constraints. With the diverse array of playback devices available, a single encoded video cannot meet all requirements. Storing multiple video copies on the server for each device would be costly. An alternative approach involves encoding the video at a high bitrate initially and then dynamically performing online trans-rating to match individual device requirements. However, it's important to note that a high bitrate also requires a larger network bandwidth for transmission. Given the variability and fluctuations in network bandwidth on the viewer's end, Streaming Service Providers (SSPs) often have to dynamically adjust the bitrate of video streams to ensure a seamless streaming experience [29]. This process of dynamically changing the bitrate of video streams to match the available network conditions is commonly referred to as Adaptive Video Streaming [30]. It allows for a more responsive and optimized viewing experience by adapting to the viewer's current network conditions. Transcoding operations fall into two primary categories: open-loop transcoding and closed-loop cascaded pixel domain transcoding. In open-loop transcoding, only the transformed coefficients in the frequency domain undergo requantization. However, during this process, the motion parameters are not re-evaluated. Consequently, there's a mismatch between the reference frames used by the encoder and decoder, leading to drift. Researchers have explored various solutions for bit rate scaling through requantization in MPEG-1/2 and H.264/AVC bitstreams [31][34]. In [31], the variable-length code words linked with the quantized discrete cosine transform (DCT) coefficients are derived from the video bitstream. These quantized coefficients undergo inverse quantization, followed by a straightforward adjustment of quantization levels to accommodate the new bitrate, a process referred to as requantization. An alternative method known as DCT coefficient dropping, which involves directly discarding high-frequency coefficients from each macroblock is proposed in [32]. Recent advancements in bit rate adaptation for H.264/AVC bitstreams are discussed in [33][34]. Authors in [33] present an efficient mixed trans-rating architecture that combines a low-complexity scheme with a drift-canceling closed-loop approach. Additionally, [34] explores a model-based trans-rating scheme integrating requantization of transform coefficients into a rate control mechanism.

- Spatial resolution reduction (Trans-sizing): Spatial resolution in the context of video refers to the physical size or dimensions of the video frame. The size of the original video stream may not always match the screen size of the viewer's device. Consequently, in order to prevent any loss of content or visual information, it is often necessary to reduce the spatial resolution of the video by either removing or combining macroblocks, a process known as downscaling. Additionally, there are situations where spatial resolution algorithms can be employed to decrease the spatial resolution without compromising the overall video quality [35]. This allows for adjustments that optimize the video's dimensions to better suit the viewing device without a significant loss in visual quality. In [36], a method was introduced for calculating downscaled motion vectors based on the original motion vectors of a video sequence. Subsequently, in [37], the process of generating motion vectors for spatial resolution reduction and the effects of refining motion estimation on this process were examined. Additionally, [38] delved into an analysis of the factors contributing to drift errors in transcoding involving spatial resolution reduction. Reducing spatial resolution becomes imperative in a transcoder to align the spatial resolution of the target video with the constraints imposed by transmission bandwidth and terminal capabilities. In [39], a transcoding technique employing merge and partition was utilized for MPEG-2 to H.264 transcoding while reducing spatial resolution. However, in MPEG-4, the Motion Estimation (ME) module incorporates four modes (Inter 16x16, Inter 8x8, SKIP, and Intra), unlike MPEG-2's single mode (Inter 16x16). Consequently, directly applying the method from [40] to MPEG4/H.264 transcoding involves handling a multitude of cases. Since the transcoding speed is primarily influenced by the complexity of the ME module in the transcoder, various efficient ME algorithms have been proposed for homogeneous transcoding and for heterogeneous video transcoding [41].

- Temporal resolution reduction: Temporal resolution reduction occurs when the viewer's device can only handle a lower frame rate compared to the original video. In such cases, the SSP needs to discard certain frames to match the viewer's device capabilities. This action can lead to issues because video frames are interrelated, and dropping frames can make motion vectors invalid for the following frames. These methods help manage the frame rate of the video to accommodate the viewer's device while mitigating potential issues caused by frame drops [42]. Temporal transcoding plays a crucial role in distributing encoded video sequences to users across channels with varying bandwidths. When transmitting video over channels with different bandwidths (e.g., in multicast sessions), the encoded video sequence must be tailored to specific bit rates for each outgoing channel. When frames are skipped, recomputing motion vectors (since the old ones become invalid for skipped frames) and prediction errors is essential. The approach involves reusing motion vectors and prediction errors from the input video sequence as much as possible. Additionally, when channel bandwidth temporarily decreases due to high user demand (subrating), adjustments are necessary. Temporal transcoding achieves this by selectively eliminating frames from the sequence, thereby reducing the frame rate without compromising the quality of non-skipped frames. To decide which frames to drop, a frame skipping policy is adopted [43].

- Video compression standard conversion: Video files are encoded using a range of video compression standards, or codecs, such as MPEG2 [44], H.264 [17], and the latest one, HEVC [45]. These compression standards are crucial because they significantly reduce the size of video files. Without them, video files would be too large to stream or store effectively using existing network and storage capacities. Most viewer's devices typically support only one or a few specific compression standards. Therefore, if the video codec used is not compatible with the viewer's device, the video needs to undergo transcoding to convert it into a format that is supported by the viewer's device [46]. This ensures that the video can be properly played back on the viewer's device.

## 3. Proposed Method

It is well known that the R-D behavior of a video is highly content-dependent. Therefore, different videos demand different coding bitrates to maintain a specific quality level, which is contingent upon the complexity of their content. The discrepancy in complexity necessitates allocating a higher bitrate to videos with greater complexity to ensure acceptable quality, in line with the principles of Rate-Distortion theory.

To expedite this content-dependent analysis for live transcoding applications and subsequently assign suitable bitrates, we employ a clustering methodology. Specifically, we utilize R-D-based clustering of varying content types and resolutions, ensuring its applicability across diverse scenarios.

Figure 3 presents an illustrative block diagram of the model formation phase of our proposed method, comprising three principal steps. The initial step involves video preparation, wherein a set of training videos undergo conversion to various resolutions and are transcoded at different bitrates to form a set of R-D curves. Subsequently, these R-D curves are subjected to clustering in the second step. Finally, the third step focuses on fitting a curve to the centroids of each cluster and identifying the knee points. Knee points are bitrates at which the R-D curve of a video coded at different resolution intersect. These points are important as they represent bitrates at which trans-sizing leads to a better quality. We also utilize our model in connection with visually lossless concept to achieve bitrate savings without introducing noticeable quality degradation.

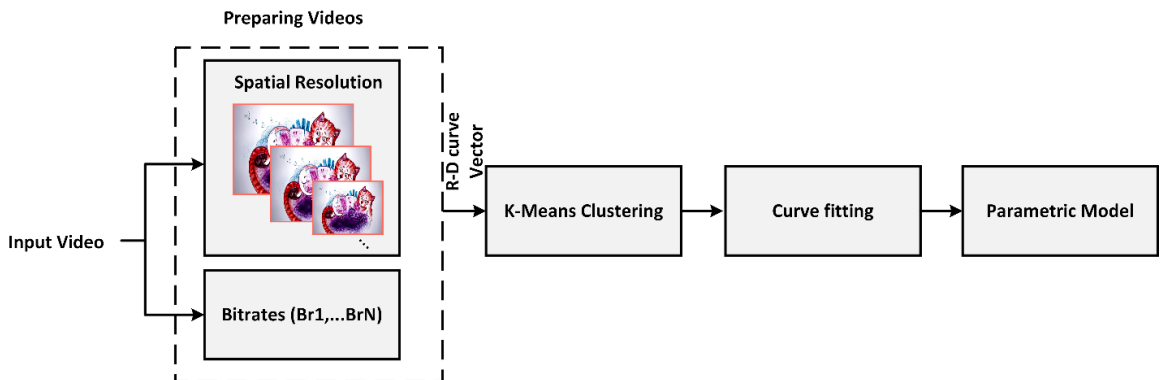

**Fig. 3** Block diagram of the model formation phase of the proposed method

### 3.1 Generation of Transcoding R-D Curves

Rate-Distortion theory is fundamentally concerned with representing a source using the fewest possible bits while maintaining a desired level of reproduction quality [47]. This trade-off between source fidelity and coding rate constitutes the essence of the rate-distortion trade-off, a pivotal consideration in the design of any lossy compression system. The R-D curve offers valuable insights into a video's behavior and enables the assignment of a bitrate that aligns with user's quality/bandwidth requirements.

To generate transcoding R-D curves, we considered a diverse set of videos sourced from the User-Generated Content (UGC) dataset, drawn from millions of YouTube videos [14]. This dataset encompasses a wide array of content types, including animations, lectures, news, TV clips, and more. We transcoded these videos across a range of bitrates. Figure 4 provides a visual representation of the R-D curves for sample video types at different bitrates from UGC. Each curve corresponds to a GOP within a video (consecutive GOPs), illustrating how, due to differences in complexity, GOPs at the same bitrate exhibit varying Peak Signal-to-Noise Ratio (PSNR) values.

In our transcoding setup, we employed a resolution of 1080p, a framerate of 30 fps, and an 8 Mbps bitrate and a GOP size of 2 seconds as the default configuration for incoming video streams (video codec is H.264). To discern the R-D behavior of diverse videos, we needed to assess the distortion at different bitrates. Consequently, our training video set underwent transcoding at bitrates ranging from 200 Kbps to 6 Mbps—a reasonable range given the uncommon nature of bitrates lower than 200 Kbps in transcoding applications. PSNR values are computed relative to the ingest video.

During the transcoding process, the transcoder has the option of changing the frame resolution (trans-sizing). To train our parametric model for resolution changes, we also altered the resolution of our training data and transcoded them. After decoding, the resolutions were restored to their native values. Figure 5 depicts the resulting R-D curves of a sample GOP for different resolutions. As evident in Figure 5, different resolutions at the same bitrates yield varying distortion values. Our transcoding parametric model, detailed in Section 3.3, is adept at modeling the impact of resolution resizing on R-D performance.

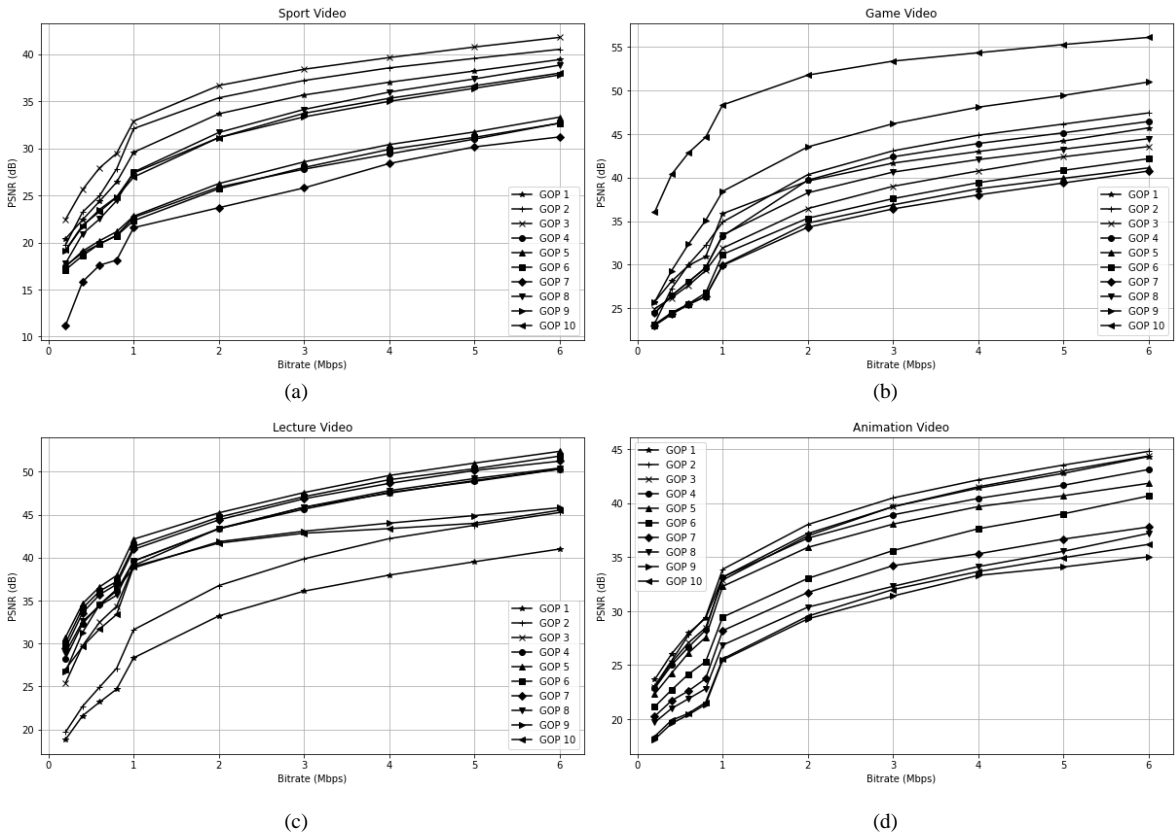

**Fig. 4** R-D curve of different video types for four consecutive GOPs, (a) Sport video, (b) Game video, (c) Lecture video, (d) Animation video

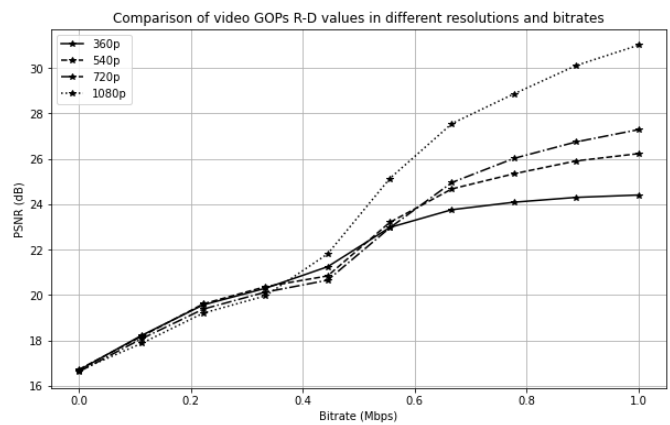

**Fig.5** The R-D curve of a sample video in different resolutions

### 3.2 R-D curve clustering

Considering the diverse transcoding R-D behavior we employ a clustering methodology to categorize videos into several clusters based on their rate-distortion characteristics, which inherently reflect their complexity levels. We employ the widely recognized K-Means [48] clustering algorithm. Through experimentation, we found that six clusters offered an effective balance between granularity and computational complexity. The input for K-Means clustering consists of the transcoding R-D curves (generated as explained in section 3.1). Figure 6 illustrates the output of K-Means clustering for all video GOPs at a resolution of 1080p, which clusters them based on their similarity. This output shows that R-D curves exhibit specific behavior, allowing them to be grouped. In Figure 7, you can observe the centroids of these clusters. Each centroid serves as the representative transcoding R-D behavior for the videos within its respective cluster. We applied a similar procedure to obtain centroids for videos at resolutions of 720p, 540p, and 360p.

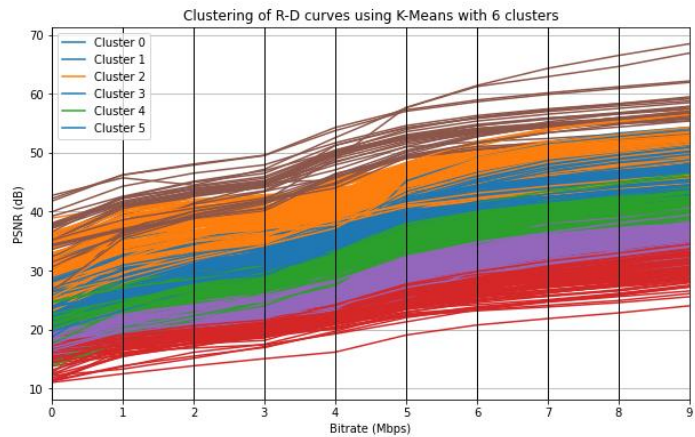

**Fig. 6** The output of K-Means clustering

### 3.3 Curve fitting

To establish a parametric model that characterizes the bitrate-quality relationship, which can subsequently be employed in R-D optimization for transcoding, we embark on a curve fitting endeavor. To calculate a suitable curve, we utilize the 10 pairs of (PSNR, bitrate) that represent the centroid of each clustering and for each resolution. In our pursuit, we explored various polynomial, spanning degrees 2 and 3, and logarithmic functions, aiming to identify the function that minimizes the mean square error (MSE) between actual values and the fitted function. Table 1 shows the average MSE for different functions in different resolutions and clusters that fitted to clusters' centroid.

**Table. 1** MSE of different functions for curve fitting

| Resolution | Linear | Polynomial | | Logarithmic (a log(bx)) |
|---|---|---|---|---|
| | | p=2 | p=3 | |
| Res 1080 | 6.370 | 1.833 | 0.460 | 0.540 |
| Res 720 | 4.438 | 1.380 | 0.432 | 0.280 |
| Res 540 | 4.325 | 1.398 | 0.443 | 0.400 |
| Res 360 | 4.200 | 1.450 | 0.452 | 0.597 |
| Average | 4.833 | 1.515 | 0.447 | 0.454 |

As we can see, polynomial functions exhibit lower MSE. Our experiments revealed that a third-order polynomial function yielded the lowest square error. We also tested higher order polynomial functions, that resulted in higher MSE due to overfitting,.

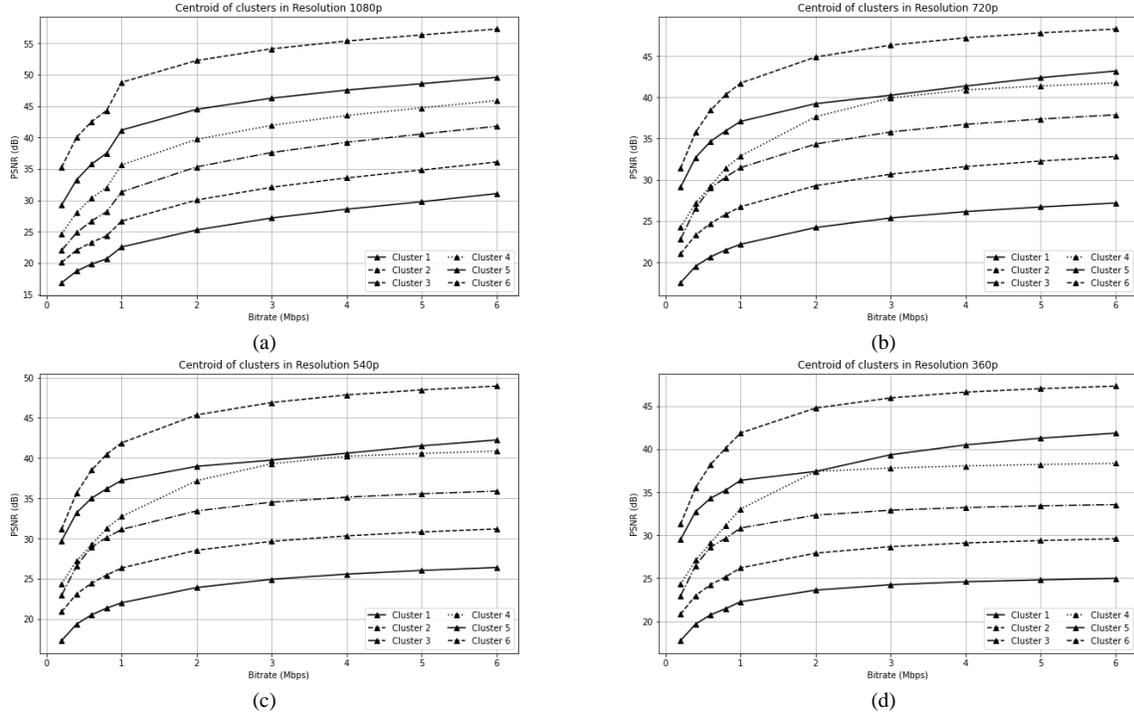

**Fig.7** The centroid of clusters in different resolutions, (a) resolution of 1080, (b) resolution of 720, (c) resolution of 540, (d) resolution of 360

Considering that we have six cluster centroids for each resolution, we need to fit six curves for each resolution. Figure 8 visually demonstrates the output of curve fitting for the 1080p resolution. Table 2 presents the respective fitted curves for each cluster centroid. Given our consideration of four different resolutions, we have four curves for each cluster centroid, each corresponding to a specific resolution. In equation in Table 2, $Q$ and $R$ represent quality (in PSNR) and bitrate values, respectively.

**Table. 2** Representing R-D of centroids using $3^{rd}$ order polynomial

| | | | | |
|---|---|---|---|---|
| Clusters | 1 | Resolution | 360p | $Q = 6.307\,R - 1.554\,R^2 + 0.129R^3 + 16.857$ |
| | | | 540p | $Q = 5.932\,R - 1.571\,R^2 + 0.133R^3 + 17.382$ |
| | | | 720p | $Q = 6.274\,R - 1.499R^2 + 0.123R^3 + 17.034$ |
| | | | 1080p | $Q = 7.627\,R - 1.643\,R^2 + 0.133R^3 + 15.749$ |
| | 2 | Resolution | 360p | $Q = 7.171\,R - 1.880\,R^2 + 0.158R^3 + 20.253$ |
| | | | 540p | $Q = 7.419\,R - 1.840\,R^2 + 0.152R^3 + 20.245$ |
| | | | 720p | $Q = 7.799R - 1.857\,R^2 + 0.152R^3 + 20.290$ |
| | | | 1080p | $Q = 9.071\,R - 1.958R^2 + 0.156R^3 + 18.596$ |
| | 3 | Resolution | 360p | $Q = 10.014\,R - 2.867\,R^2 + 0.252R^3 + 22.660$ |
| | | | 540p | $Q = 10.715\,R - 2.904R^2 + 0.250R^3 + 22.494$ |
| | | | 720p | $Q = 11.467\,R - 2.987\,R^2 + 0.254R^3 + 22.166$ |
| | | | 1080p | $Q = 12.650\,R - 2.915R^2 + 0.236R^3 + 20.103$ |
| | 4 | Resolution | 360p | $Q = 13.527\,R - 3.578\,R^2 + 0.297R^3 + 22.210$ |
| | | | 540p | $Q = 12.318\,R - 2.880R^2 + 0.223R^3 + 22.658$ |
| | | | 720p | $Q = 12.845\,R - 2.985R^2 + 0.231R^3 + 22.410$ |
| | | | 1080p | $Q = 14.786\,R - 3.571\,R^2 + 0.294R^3 + 22.564$ |
| | 5 | Resolution | 360p | $Q = 7.619\,R - 1.888\,R^2 + 0.161R^3 + 29.598$ |
| | | | 540p | $Q = 9.405R - 2.604R^2 + 0.234R^3 + 29.483$ |
| | | | 720p | $Q = 10.075\,R - 2.700R^2 + 0.239R^3 + 28.743$ |
| | | | 1080p | $Q = 15.563\,R - 4.010\,R^2 + 0.341R^3 + 27.468$ |
| | 6 | Resolution | 360p | $Q = 14.048\,R - 3.814R^2 + 0.327R^3 + 30.278$ |
| | | | 540p | $Q = 14.374R - 3.801R^2 + 0.323R^3 + 30.229$ |
| | | | 720p | $Q = 13.695\,R - 3.644R^2 + 0.311R^3 + 30.558$ |
| | | | 1080p | $Q = 17.415R - 4.521R^2 + 0.383R^3 + 33.335$ |

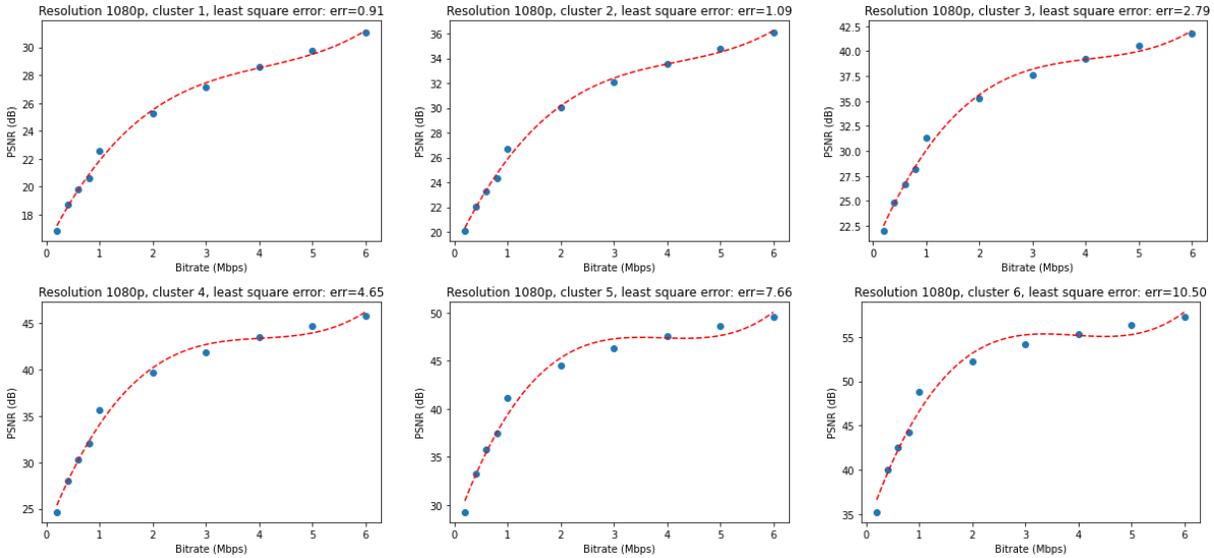

**Fig. 8** The curve fits for different clusters in resolution of 1080p

### 3.4 Knee points and trans-sizing

Now we turn our attention to the behavior of cluster centroids across varying resolutions. It became evident through our experiments that in a number of clusters (corresponding to certain video types), the R-D curves for different resolutions intersect with each other. These intersection points (or knee points) hold significance, as they indicate the bitrates at which transitioning to a different resolution during transcoding can offer advantages in an R-D context. Figure 9 offers a consolidated view of the fitted curves for different resolutions within each cluster.

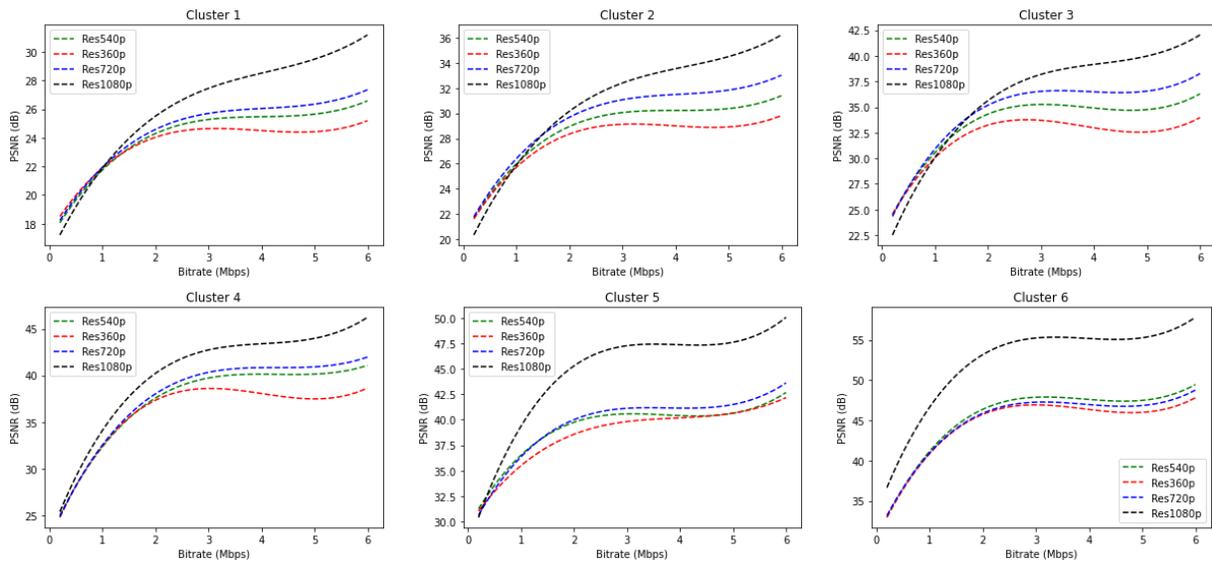

**Fig. 9** Consolidation of cluster centroid fitted curves for differernt resolutions in each cluster

As Figure 9 illustrates, for videos in Cluster 1, lower resolutions provide superior quality at lower bitrates. However, for videos with lower complexity (e.g., Clusters 5 and 6, where at the same bitrate, higher PSNR indicates relatively simple content), the higher resolution (1080p) consistently delivers better quality across all bitrates. Consequently, intersection points (knee points) in various clusters guide our decisions regarding trans-sizing, enabling us to select the optimal resolution based on the transcoding's target bit rate.

We identify these intersections by solving the equations derived from equating the polynomial functions for pairs of curves. Consequently, for each cluster, we establish conditional equations that ascertain the best resolution (in an R-D sense) based on the bitrate. Equations (1 to 6) presented below represent the conditional equations for Clusters 1 through 6.

Using these equations, for an ingest video to the transcoder and based on the target bitrate the best resolution can be determined. For example, if an ingest video is identified to belong to cluster 1, and if the target bitrate is $2\ Mbps$ then this video should be resized to 360 first and then encoded. This will guarantee the best possible quality (in PSNR term).

$$cluster1 \rightarrow \begin{cases} Res360p & R \leq 0.876 \\ Res720p & 0.876 < R \leq 1.061 \\ Res1080p & R > 1.061 \end{cases} \tag{1}$$

$$cluster\ 2 \rightarrow \begin{cases} Res1080p & R > 1.499 \\ Res\ 720p & o.w. \end{cases} \tag{2}$$

$$cluster\ 3 \rightarrow \begin{cases} Res1080p & R > 1.647 \\ Res720p & 0.349 < R \leq 1.647 \\ Res540p & 0.239 < R \leq 0.349 \\ Res360p & R \leq 0.239 \end{cases} \tag{3}$$

$$cluster\ 4 \rightarrow \begin{cases} Res540p & R \leq 0.038 \\ Res720p & 0.038 < R \leq 0.049 \\ Res1080p & R > 0.049 \end{cases} \tag{4}$$

$$cluster\ 5 \rightarrow \begin{cases} Res1080p & R > 0.355 \\ Res\ 540p & o.w. \end{cases} \tag{5}$$

$$cluster\ 6 \rightarrow Res1080p \tag{6}$$

where $R$ and $Res$ are the target transcoding bitrate and resolution respectively.

### 3.5 Bitrate adjustment based on visually lossless and near-zero-slope

As mentioned in [49], compressed videos with a PSNR value of 40 dB or above, typically constituted visually lossless coding. Therefore, if based on our model the output video of transcoding yields a PSNR that is more than 40 dB, we can reduce the bitrate to a level that generate a PSNR of 40 dB. As can be seen in Fig. 9, in clusters 3,4,5, and 6, this condition can be applied and the target bitrate can be decreased by more than 3 Mbps. Equations (7) to (12) show the bitrate corresponding to a 40 dB quality in each cluster. Therefore, if the transcoding target bit rate exceed the bitrates identified in Equations (7 to 12), we can lower the target bit rate to the threshold bitrate for the cluster that that video belongs to.

$$cluster1 \rightarrow \begin{cases} Res360p & R \geq 8.808 \\ Res540p & R \geq 8.951 \\ Res720p & R \geq 8.802 \\ Res1080p & R \geq 8.041 \end{cases} \tag{7}$$

$$cluster\ 2 \rightarrow \begin{cases} Res360p & R \geq 8.229 \\ Res540p & R \geq 8.046 \\ Res720p & R \geq 7.757 \\ Res1080p & R \geq 7.072 \end{cases} \tag{8}$$

$$cluster\ 3 \rightarrow \begin{cases} Res360p & R \geq 7.175 \\ Res540p & R \geq 8.95 \\ Res720p & R \geq 6.445 \\ Res1080p & R \geq 5.018 \end{cases} \tag{9}$$

$$cluster\ 4 \rightarrow \begin{cases} Res360p & R \geq 6.379 \\ Res540p & R \geq 3.377 \\ Res720p & R \geq 2.772 \\ Res1080p & R \geq 1.950 \end{cases} \tag{10}$$

$$cluster\ 5 \rightarrow \begin{cases} Res360p & R \geq 3.385 \\ Res540p & R \geq 2.155 \\ Res720p & R \geq 1.998 \\ Res1080p & R \geq 1.077 \end{cases} \tag{11}$$

$$cluster\ 6 \rightarrow \begin{cases} Res360p & R \geq 0.891 \\ Res540p & R \geq 0.862 \\ Res720p & R \geq 0.880 \\ Res1080p & R \geq 0.429 \end{cases} \tag{12}$$

We also can consider the slope of R-D curves in each cluster and decrease the bitrate values in regions where R-D

slope is near zero. Figure. 10 shows the 1080 resolution curve in cluster 5 and its first derivative. In this diagram, the region with the near zero slope (less than 0.1) is highlighted. As we can see in Fig. 10, for a video in this cluster and for a transcoding target bitrate falling in this interval the target bit rate can be reduced to around 3 Mbps without considerable changes in video quality. The near-zero interval can be adjusted based on the acceptable "near" zero slope range.

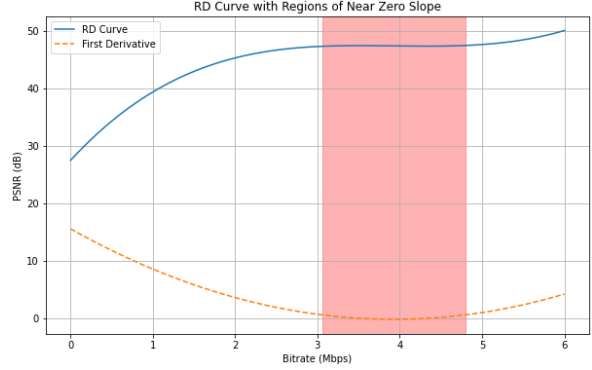

Fig. 10 R-D curve with regions of near zero slope for cluster 5

Using each of the equations in Table 2, we can identify the near zero slope region based on the first derivative of the equations. Equation (13 to 18) shows the near-zero-interval in each cluster based on a less that 0.1 threshold for the slope. For some cluster/resolution combination this region does not exist, and we have to maintain the original target transcoding bitrate.

$$cluster\,1 \rightarrow \begin{cases} Res360p & 3.724 \leq R \leq 4.306 \\ Res540p & 3.003 \leq R \leq 4.875 \\ Res720p & - \\ Res1080p & - \end{cases} \tag{13}$$

$$cluster\,2 \rightarrow \begin{cases} Res360p & 3.073 \leq R \leq 4.865 \\ Res540p & 3.564 \leq R \leq 4.515 \\ Res1080p & - \\ Res1080p & - \end{cases} \tag{14}$$

$$cluster\,3 \rightarrow \begin{cases} Res360p & 2.673 \leq R \leq 4.915 \\ Res540p & 2.963 \leq R \leq 4.785 \\ Res720p & 3.253 \leq R \leq 4.585 \\ Res1080p & - \end{cases} \tag{15}$$

$$cluster\,4 \rightarrow \begin{cases} Res360p & 2.993 \leq R \leq 5.035 \\ Res540p & 3.794 \leq R \leq 4.815 \\ Res720p & 3.914 \leq R \leq 4.705 \\ Res1080p & - \end{cases} \tag{16}$$

$$cluster\,5 \rightarrow \begin{cases} Res360p & - \\ Res540p & 3.003 \leq R \leq 4.414 \\ Res720p & 3.253 \leq R \leq 4.274 \\ Res1080p & 3.423 \leq R \leq 4.414 \end{cases} \tag{17}$$

$$cluster\,6 \rightarrow \begin{cases} Res360p & 2.943 \leq R \leq 4.835 \\ Res540p & 3.113 \leq R \leq 4.725 \\ Res720p & 3.083 \leq R \leq 4.725 \\ Res1080p & 3.293 \leq R \leq 4.575 \end{cases} \tag{18}$$

## 4. Experimental Results

We implemented our proposed method, on an Intel(R) Core (TM) i7-3770 CPU running at 3.40 GHz. The offline phase involved utilizing the UGC dataset [14] as our "training" dataset for constructing the R-D curves at different resolutions. This dataset comprised 1,500 video clips spanning 15 categories. Each clip had a duration of 20 seconds. In our selection process, we focused on 13 available video types from the UGC dataset, including Animation, Game, Lecture, Sport, among others in resolution 1080, and opted for 15 videos from each category for training purposes. Notably, each video had a length of 20 seconds, and we considered GOP segments with a duration of 2 seconds. In our case, we used FFmpeg to generate 2-second shots from the YouTube UGC dataset, resulting in a total of 1,950 shots. These shots were then transcoded using AVC/H264. Furthermore, we calculated PSNR metric. We generated other resolutions, 360p, 540p, and 720p, from the 1080p videos using FFmpeg.

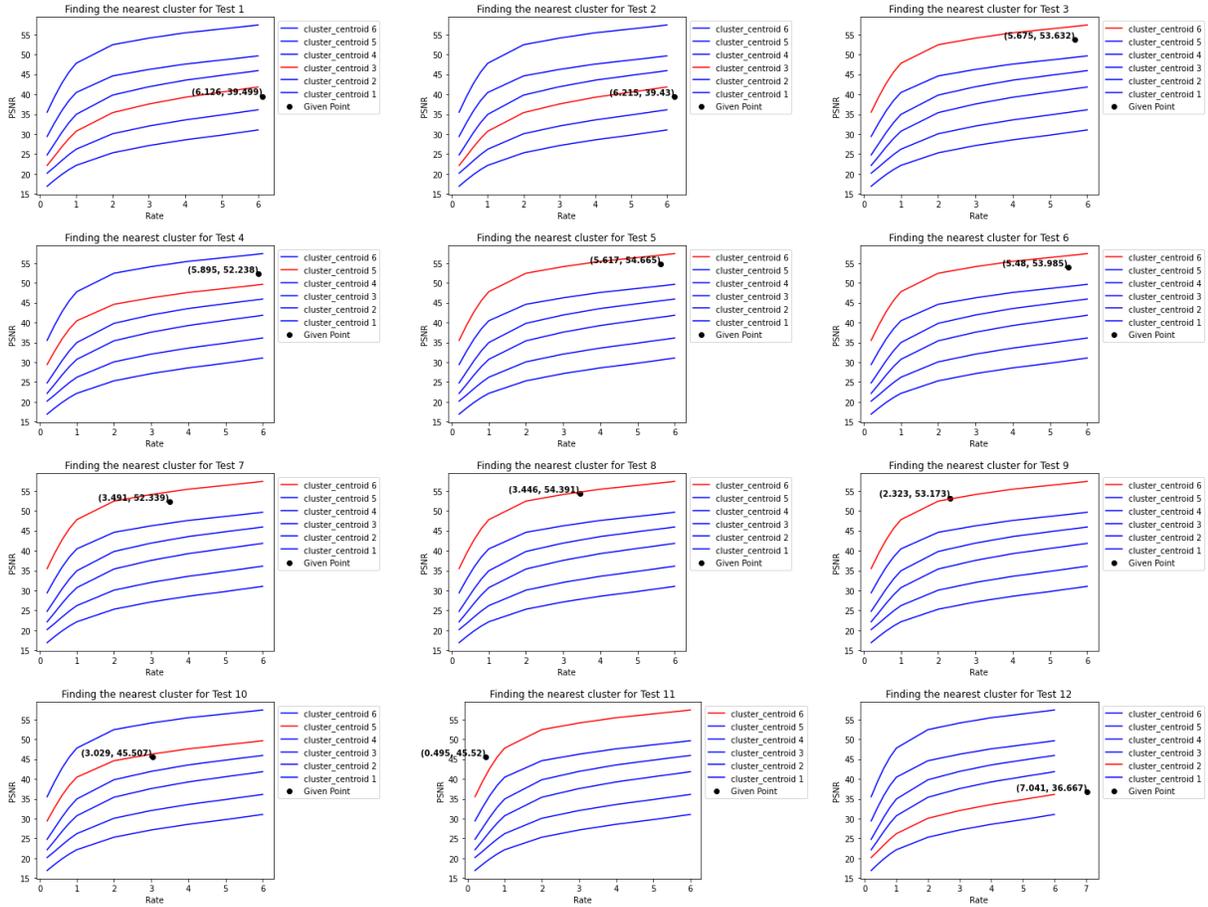

**Fig.11** Selected centroid based on Eculdian distance for differnet input test videos

The characteristics of the UGC dataset and our selected videos are summarized in Table 3:

**Table 3.** Characteristics of UGC dataset and our selected videos

|  | YouTube | Our selected videos |
|---|---|---|
| Number of clips | 1500 | 195 |
| Number of categories | 15 | 13 |
| Duration of clips | 20s | 20s |
| Resolution | 360p, 480p, 720p, 1080p, 4K | 1080p |

### 4.1 Transcoding R-D behavior prediction

We now show how our framework can be utilized for transcoding R-D behavior prediction. We tested our approach on 10% of our shots that is around 200 videos. Different video segments were coded at various bitrates all at a resolution of 1080p with 20 second lengths. Our first task was to determine the relevant cluster for this input video. To achieve this, we relied on the PSNR and bitrate values for the ingest video. By calculating the Euclidean distance between the PSNR of the input video and all centroids at the same bitrate (using equations in Table 2 calculated at that bitrate), we identified the nearest centroid. We could assign the correct cluster with around 82.17% accuracy on our test data. Table 4 shows the bitrate and PSNR for a random subset of the selected test videos (12 videos from 200 videos). Figure 11 graphically depicts the selected centroids for each input video. The centroid with the lowest distance is highlighted in red. This selection determines the cluster to which ingest video belongs. We then assign the R-D curve of this cluster centroid as the R-D curve of the ingest video, enabling us to predict the transcoding R-D behavior of the video at different bitrates and resolutions. This information is invaluable for making quality/bitrate optimization decisions. Therefore, our approach eliminates the need for exhaustively coding a video at different bitrates and different resolutions making our approach suitable for live video streaming.

**Table 4.** Test videos in resolution 1080p

| Video name | Bitrate (Mbps) | PSNR (dB) |
|---|---|---|
| Test1 | 6.127 | 39.499 |
| Test2 | 6.215 | 39.430 |
| Test3 | 5.675 | 53.632 |
| Test4 | 5.895 | 52.238 |
| Test5 | 5.617 | 54.665 |
| Test6 | 5.480 | 53.985 |
| Test7 | 3.491 | 52.339 |
| Test8 | 3.446 | 54.391 |
| Test9 | 2.323 | 53.173 |
| Test10 | 3.029 | 45.507 |
| Test11 | 0.494 | 45.516 |
| Test12 | 7.041 | 36.667 |

In the subsequent step, we compared the predicted R-D curve with the actual R-D curve of the input videos. To achieve this, we transcoded the input videos at different bitrates, ranging from 0.2 Mbps to 6 Mbps, and afterwards calculated the distortion for each of these transcoded versions. We then associated the resulting R-D values with the nearest centroid cluster, demonstrating the accuracy of our predictions. Figure 12 illustrates the R-D curve for the input video alongside the nearest centroid.

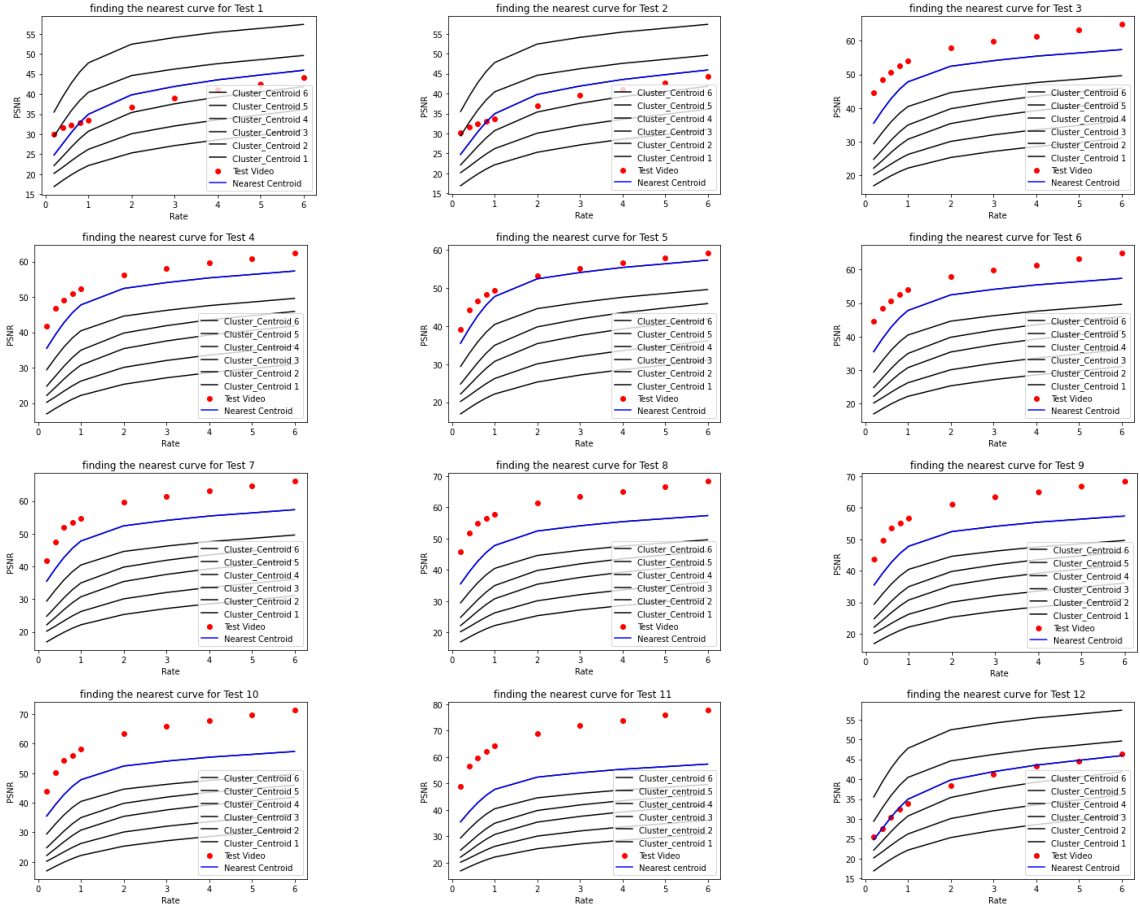

**Fig. 12** Finding the nearest centorid to input video R-D curve

Comparing Figure 11 and Figure 12, it becomes evident that our predictions align closely with the actual R-D curves.

### 4.2 Trans-sizing optimization

By assigning the transcoder's incoming video segment to a cluster, the R-D behavior of that video segment at different bitrates and different resolutions can be predicted without having to encode the video. For example, our test video 12 from Table 4 is determined to belong to Cluster 2. Therefore, based on Equation (2), if the transcoding target bitrate is less than 1.49 Mbps, it is better (in R-D terms) to resize this video to 720p while transcoding. Thus, our model can be used for trans-sizing optimization. It should be mentioned that after decoding, the frames are restored to their native resolution using FFmpeg's interpolation function. To evaluate the effectiveness of the proposed method, we randomly selected 11 videos from different types as test videos to the transcoder. Assuming each video is 20 seconds in length and contains 10 GOPs, encoded at 8Mbps bitrate. Based on this bitrate and PSNR, the nearest cluster for each GOP is determined. The GOPs for different videos and the identified clusters for each of them are presented in Table 5.

**Table. 5** Input test videos and identified clusters for each GOP

| Input Video | GOPS | | | | | | | | | |
|---|---|---|---|---|---|---|---|---|---|---|
| | GOP1 | GOP2 | GOP3 | GOP4 | GOP5 | GOP6 | GOP7 | GOP8 | GOP9 | GOP10 |
| Test_1 | Cluster 3 | Cluster 3 | Cluster 3 | Cluster 3 | Cluster 3 | Cluster 3 | Cluster 3 | Cluster 3 | Cluster 3 | Cluster 3 |
| Test_2 | Cluster 6 | Cluster 6 | Cluster 6 | Cluster 6 | Cluster 6 | Cluster 4 | Cluster 1 | Cluster 1 | Cluster 1 | Cluster 1 |
| Test_3 | Cluster 5 | Cluster 5 | Cluster 5 | Cluster 4 | Cluster 3 | Cluster 2 | Cluster 2 | Cluster 2 | Cluster 2 | Cluster 2 |
| Test_4 | Cluster 3 | Cluster 3 | Cluster 3 | Cluster 3 | Cluster 3 | Cluster 3 | Cluster 3 | Cluster 3 | Cluster 3 | Cluster 3 |
| Test_5 | Cluster 6 | Cluster 6 | Cluster 6 | Cluster 6 | Cluster 6 | Cluster 6 | Cluster 6 | Cluster 6 | Cluster 6 | Cluster 6 |
| Test_6 | Cluster 5 | Cluster 5 | Cluster 4 | Cluster 3 | Cluster 3 | Cluster 3 | Cluster 3 | Cluster 3 | Cluster 4 | Cluster 6 |
| Test_7 | Cluster 4 | Cluster 4 | Cluster 4 | Cluster 4 | Cluster 4 | Cluster 4 | Cluster 4 | Cluster 4 | Cluster 4 | Cluster 4 |
| Test_8 | Cluster 2 | Cluster 2 | Cluster 2 | Cluster 2 | Cluster 2 | Cluster 2 | Cluster 3 | Cluster 2 | Cluster 2 | Cluster 1 |
| Test_9 | Cluster 1 | Cluster 1 | Cluster 1 | Cluster 1 | Cluster 1 | Cluster 1 | Cluster 1 | Cluster 1 | Cluster 1 | Cluster 1 |
| Test_10 | Cluster 4 | Cluster 4 | Cluster 4 | Cluster 5 | Cluster 4 | Cluster 4 | Cluster 4 | Cluster 4 | Cluster 5 | Cluster 5 |
| Test_11 | Cluster 1 | Cluster 1 | Cluster 2 | Cluster 4 | Cluster 4 | Cluster 2 | Cluster 1 | Cluster 1 | Cluster 1 | Cluster 1 |

After finding the related clusters based on our parametric model and the target bitrate (selected from a wide range of typical bitrates used in video streaming applications), we can choose the suitable resolution to achieve the best quality. Table 6 shows the selected target bitrate for each test video. By considering the target bitrate and the corresponding cluster from Table 5, we select suitable resolutions for each GOP based on Equations (1) to (6). The last column of Table 6 indicates the average PSNR for each test video with trans-sizing using based on prediction from our model and without trans-sizing and coding the video in its native resolution. As we can see, by applying our model, the final quality is either better or equal.

**Table. 6** Input test videos and suggested resolution for each GOP

| Input Video | Target transcoding bitrate (Mbps) | GOPS | | | | | | | | | | Average PSNR | | $\Delta PSNR$ (dB) |
|---|---|---|---|---|---|---|---|---|---|---|---|---|---|---|
| | | GOP1 | GOP2 | GOP3 | GOP4 | GOP5 | GOP6 | GOP7 | GOP8 | GOP9 | GOP10 | With Trans sizing | Without Trans sizing | |
| Test_1 | 1 | Res 720 | Res 720 | Res 720 | Res 720 | Res 720 | Res 720 | Res 720 | Res 720 | Res 720 | Res 720 | **30.90** | 30.07 | 0.83 |
| Test_2 | 0.8 | Res 1080 | Res 1080 | Res 1080 | Res 1080 | Res 1080 | Res 1080 | Res 360 | Res 360 | Res 360 | Res 360 | **33.89** | 33.85 | 0.04 |
| Test_3 | 0.3 | Res 540 | Res 540 | Res 540 | Res 1080 | Res 720 | Res 720 | Res 720 | Res 720 | Res 720 | Res 720 | **26.07** | 25.14 | 0.93 |
| Test_4 | 0.2 | Res 360 | Res 360 | Res 360 | Res 360 | Res 360 | Res 360 | Res 360 | Res 360 | Res 360 | Res 360 | **24.77** | 22.52 | 2.25 |
| Test_5 | 2 | Res 1080 | Res 1080 | Res 1080 | Res 1080 | Res 1080 | Res 1080 | Res 1080 | Res 1080 | Res 1080 | Res 1080 | 53.15 | 53.15 | 0.00 |
| Test_6 | 0.5 | Res 1080 | Res 1080 | Res 1080 | Res 720 | Res 720 | Res 720 | Res 720 | Res 720 | Res 1080 | Res 1080 | **30.36** | 29.63 | 0.73 |
| Test_7 | 1 | Res 1080 | Res 1080 | Res 1080 | Res 1080 | Res 1080 | Res 1080 | Res 1080 | Res 1080 | Res 1080 | Res 1080 | 34.07 | 34.07 | 0.00 |
| Test_8 | 1 | Res 720 | Res 720 | Res 720 | Res 720 | Res 720 | Res 720 | Res 720 | Res 720 | Res 720 | Res 720 | **26.84** | 26.28 | 0.56 |
| Test_9 | 0.7 | Res 360 | Res 360 | Res 360 | Res 360 | Res 360 | Res 360 | Res 360 | Res 360 | Res 360 | Res 360 | **20.55** | 20.32 | 0.23 |
| Test_10 | 0.3 | Res 1080 | Res 1080 | Res 1080 | Res 540 | Res 1080 | Res 1080 | Res 1080 | Res 1080 | Res 540 | Res 540 | **28.30** | 28.22 | 0.08 |
| Test_11 | 1 | Res 720 | Res 720 | Res 1080 | Res 1080 | Res 1080 | Res 720 | Res 720 | Res 720 | Res 720 | Res 720 | **24.98** | 24.79 | 0.19 |

### 4.3 Bitrate saving based on visually lossless and near zero-slop region

In section 3.5 we introduced two methods for bitrate saving. One is to decrease the bitrate until the PSNR reached 40 dB which is the minimum for visually lossless quality. Table 7 displays the assigned bitrate for each GOP of the test videos from Table 5 based on this assumption. When considering a target transcoded bitrate for each test video, our model dynamically proposes a lower bitrate that achieves the desired 40 dB PSNR without compromising quality. However, if the related cluster does not exceed the 40 dB threshold, our model maintains the original target bitrate. Our proposed method reduces the assigned bitrate during transcoding, leading to improved bandwidth efficiency while maintaining perceptual quality. As can be seen from Table 7, bitrate savings up to 46% can be achieved using our model.

Table. 7 Input test videos and assigned bitrate for each GOP

| | | Test_1 | Test_2 | Test_3 | Test_4 | Test_5 | Test_6 | Test_7 | Test_8 | Test_9 | Test_10 | Test_11 |
|---|---|---|---|---|---|---|---|---|---|---|---|---|
| GOP1 | Target bitrate | 6.000 | 3.000 | 3.000 | 6.000 | 1.000 | 2.000 | 3.000 | 5.000 | 3.000 | 3.000 | 3.000 |
| | Model proposed bitrate | 5.018 | 0.428 | 1.077 | 5.018 | 0.428 | 1.077 | 1.950 | 5.000 | 3.000 | 1.950 | 3.000 |
| GOP2 | Target bitrate | 6.000 | 3.000 | 3.000 | 6.000 | 1.000 | 2.000 | 3.000 | 5.000 | 3.000 | 3.000 | 3.000 |
| | Model proposed bitrate | 5.018 | 0.428 | 1.077 | 5.018 | 0.428 | 1.077 | 1.950 | 5.000 | 3.000 | 1.950 | 3.000 |
| GOP3 | Target bitrate | 6.000 | 3.000 | 3.000 | 6.000 | 1.000 | 2.000 | 3.000 | 5.000 | 3.000 | 3.000 | 3.000 |
| | Model proposed bitrate | 5.018 | 0.428 | 1.077 | 5.018 | 0.428 | 1.950 | 1.950 | 5.000 | 3.000 | 1.950 | 3.000 |
| GOP4 | Target bitrate | 6.000 | 3.000 | 3.000 | 6.000 | 1.000 | 2.000 | 3.000 | 5.000 | 3.000 | 3.000 | 3.000 |
| | Model proposed bitrate | 5.018 | 0.428 | 1.950 | 5.018 | 0.428 | 2.000 | 1.950 | 5.000 | 3.000 | 1.077 | 1.950 |
| GOP5 | Target bitrate | 6.000 | 3.000 | 3.000 | 6.000 | 1.000 | 2.000 | 3.000 | 5.000 | 3.000 | 3.000 | 3.000 |
| | Model proposed bitrate | 5.018 | 0.428 | 3.000 | 5.018 | 0.428 | 2.000 | 1.950 | 5.000 | 3.000 | 1.950 | 1.950 |
| GOP6 | Target bitrate | 6.000 | 3.000 | 3.000 | 6.000 | 1.000 | 2.000 | 3.000 | 5.000 | 3.000 | 3.000 | 3.000 |
| | Model proposed bitrate | 5.018 | 1.950 | 3.000 | 5.018 | 0.428 | 2.000 | 1.950 | 5.000 | 3.000 | 1.950 | 3.000 |
| GOP7 | Target bitrate | 6.000 | 3.000 | 3.000 | 6.000 | 1.000 | 2.000 | 3.000 | 5.000 | 3.000 | 3.000 | 3.000 |
| | Model proposed bitrate | 5.018 | 3.000 | 3.000 | 5.018 | 0.428 | 2.000 | 1.950 | 5.000 | 3.000 | 1.950 | 3.000 |
| GOP8 | Target bitrate | 6.000 | 3.000 | 3.000 | 6.000 | 1.000 | 2.000 | 3.000 | 5.000 | 3.000 | 3.000 | 3.000 |
| | Model proposed bitrate | 5.018 | 3.000 | 3.000 | 5.018 | 0.428 | 2.000 | 1.950 | 5.000 | 3.000 | 1.950 | 3.000 |
| GOP9 | Target bitrate | 6.000 | 3.000 | 3.000 | 6.000 | 1.000 | 2.000 | 3.000 | 5.000 | 3.000 | 3.000 | 3.000 |
| | Model proposed bitrate | 5.018 | 3.000 | 3.000 | 5.018 | 0.428 | 1.950 | 1.950 | 5.000 | 3.000 | 1.077 | 3.000 |
| GOP10 | Target bitrate | 6.000 | 3.000 | 3.000 | 6.000 | 1.000 | 2.000 | 3.000 | 5.000 | 3.000 | 3.000 | 3.000 |
| | Model proposed bitrate | 5.018 | 3.000 | 3.000 | 5.018 | 0.428 | 0.428 | 1.950 | 5.000 | 3.000 | 1.077 | 3.000 |
| Total bitrate using our model | | 50.18 | 16.09 | 23.18 | 50.18 | 4.28 | 16.48 | 19.50 | 50.00 | 30.00 | 16.88 | 28.95 |
| Total bitrate based on target transcoding bitrate | | 60.00 | 30.00 | 30.00 | 60.00 | 10.00 | 20.00 | 30.00 | 50.00 | 30.00 | 30.00 | 30.00 |
| Bitrate Saving % | | 16.36 | 46.36 | 22.73 | 16.36 | 57.20 | 17.60 | 35.00 | 0.00 | 0.00 | 43.73 | 3.50 |

Our model can also be used to identify the near-zero-slope interval in each cluster which can result in bit saving with minimal visual impact. The saving can be achieved by decreasing the rate to the lower limit of the near-zero-slope interval without sensible change in quality. Table 8 shows the assigned bitrate based on this approach. For each GOP after determining the GOP's clusters based on Equations (13) to (18), if applicable, we reduce the bitrate to the lower limit of the near zero slope. Table 8 shows the results for resolution 1080p. As clear from the values in Table 8, bitrate savings up to 28% can be achieved using our model.

Table. 8 Input test videos and assigned bitrate for each GOP

| | | Test_1 | Test_2 | Test_3 | Test_4 | Test_5 | Test_6 | Test_7 | Test_8 | Test_9 | Test_10 | Test_11 |
|---|---|---|---|---|---|---|---|---|---|---|---|---|
| GOP1 | Target bitrate | 5.020 | 4.575 | 4.414 | 5.000 | 4.575 | 4.414 | 3.000 | 3.000 | 3.000 | 4.000 | 4.000 |
| | Model proposed bitrate | 5.020 | 3.293 | 3.423 | 5.000 | 3.293 | 3.423 | 3.000 | 3.000 | 3.000 | 4.000 | 4.000 |
| GOP2 | Target bitrate | 5.020 | 4.575 | 4.414 | 5.000 | 4.575 | 4.414 | 3.000 | 3.000 | 3.000 | 4.000 | 4.000 |
| | Model proposed bitrate | 5.020 | 3.293 | 3.423 | 5.000 | 3.293 | 3.423 | 3.000 | 3.000 | 3.000 | 4.000 | 4.000 |
| GOP3 | Target bitrate | 5.020 | 4.575 | 4.414 | 5.000 | 4.575 | 4.414 | 3.000 | 3.000 | 3.000 | 4.000 | 4.000 |
| | Model proposed bitrate | 5.020 | 3.293 | 3.423 | 5.000 | 3.293 | 4.414 | 3.000 | 3.000 | 3.000 | 4.000 | 4.000 |
| GOP4 | Target bitrate | 5.020 | 4.575 | 4.414 | 5.000 | 4.575 | 4.414 | 3.000 | 3.000 | 3.000 | 4.000 | 4.000 |
| | Model proposed bitrate | 5.020 | 3.293 | 4.414 | 5.000 | 3.293 | 4.414 | 3.000 | 3.000 | 3.000 | 3.423 | 4.000 |
| GOP5 | Target bitrate | 5.020 | 4.575 | 4.414 | 5.000 | 4.575 | 4.414 | 3.000 | 3.000 | 3.000 | 4.000 | 4.000 |
| | Model proposed bitrate | 5.020 | 3.293 | 4.414 | 5.000 | 3.293 | 4.414 | 3.000 | 3.000 | 3.000 | 4.000 | 4.000 |
| GOP6 | Target bitrate | 5.020 | 4.575 | 4.414 | 5.000 | 4.575 | 4.414 | 3.000 | 3.000 | 3.000 | 4.000 | 4.000 |
| | Model proposed bitrate | 5.020 | 4.575 | 4.414 | 5.000 | 3.293 | 4.414 | 3.000 | 3.000 | 3.000 | 4.000 | 4.000 |
| GOP7 | Target bitrate | 5.020 | 4.575 | 4.414 | 5.000 | 4.575 | 4.414 | 3.000 | 3.000 | 3.000 | 4.000 | 4.000 |
| | Model proposed bitrate | 5.020 | 4.575 | 4.414 | 5.000 | 3.293 | 4.414 | 3.000 | 3.000 | 3.000 | 4.000 | 4.000 |
| GOP8 | Target bitrate | 5.020 | 4.575 | 4.414 | 5.000 | 4.575 | 4.414 | 3.000 | 3.000 | 3.000 | 4.000 | 4.000 |
| | Model proposed bitrate | 5.020 | 4.575 | 4.414 | 5.000 | 3.293 | 4.414 | 3.000 | 3.000 | 3.000 | 4.000 | 4.000 |
| GOP9 | Target bitrate | 5.020 | 4.575 | 4.414 | 5.000 | 4.575 | 4.414 | 3.000 | 3.000 | 3.000 | 4.000 | 4.000 |
| | Model proposed bitrate | 5.020 | 4.575 | 4.414 | 5.000 | 3.293 | 4.414 | 3.000 | 3.000 | 3.000 | 3.423 | 4.000 |
| GOP10 | Target bitrate | 5.020 | 4.575 | 4.414 | 5.000 | 4.575 | 4.414 | 3.000 | 3.000 | 3.000 | 4.000 | 4.000 |
| | Model proposed bitrate | 5.020 | 4.575 | 4.414 | 5.000 | 3.293 | 3.293 | 3.000 | 3.000 | 3.000 | 3.423 | 4.000 |
| Total bitrate using our model | | 50.200 | 45.750 | 44.140 | 50.000 | 45.750 | 44.140 | 30.000 | 30.000 | 30.000 | 40.000 | 40.000 |
| Total bitrate based on target transcoding bitrate | | 50.020 | 39.340 | 41.167 | 50.000 | 32.930 | 41.037 | 30.000 | 30.000 | 30.000 | 38.269 | 40.000 |
| Bitrate Saving % | | 0.000 | 14.011 | 6.735 | 0.000 | 28.022 | 7.029 | 0.000 | 0.000 | 0.000 | 4.328 | 0.000 |

## 5. Conclusion

In this paper, we propose a parametric model designed specifically for video transcoding. As streaming video service providers strive to optimize bitrate while ensuring user satisfaction with delivered content quality, they face a significant challenge. The diverse array of user devices, varying resolutions, and fluctuating Internet speeds necessitate careful consideration during the transcoding process. Our proposed parametric model finds practical application in live streaming scenarios, where it provides information to dynamically adjusts bitrate and resolution. Notably, this model possesses the capability to predict the rate-distortion (R-D) behavior of the transcoder's ingest video across different bitrates and resolutions. We also showed how our model can be used to introduce better quality via trans-sizing or bit rate savings by identifying visually loss less or near-zero-slope RD intervals. Looking ahead, further research in this area holds promise. Fine-tuning the model and exploring advanced feature extraction and classification methods could yield even more precise and effective transcoding solutions. One potential avenue for future investigation involves feature extraction from input videos, allowing us to identify relevant clusters using classification techniques and thereby enhance prediction accuracy.

# REFERENCES

[1] Cisco Visual Networking Index: Forecast and Trends White Paper 2017–2022, San Jose, CA, USA, vol. 1, 2017.

[2] K. Zhu, C. Li, V. Asari, and D. Saupe, "No-reference video quality assessment based on artifact measurement and statistical analysis," *IEEE Transactions on Circuits and Systems for Video Technology*, vol. 25, no. 4, pp. 533–546, Apr. 2014.

[3] W. Zou, F. Yang, J. Song, S. Wan, W. Zhang, and H. R. Wu, "Eventbased perceptual quality assessment for HTTP-based video streaming with playback interruption," *IEEE Transactions on Multimedia*, vol. 20, no. 6, pp. 1475–1488, 2017.

[4] D. Ghadiyaram, J. Pan, and A. C. Bovik, "A subjective and objective study of stalling events in mobile streaming videos," *IEEE Transactions on Circuits and Systems for Video Technology*, vol. 29, no. 1, pp. 183–197, 2017.

[5] G. I. P. Report, "https://www.sandvine.com/trends/global-internet-phenomena/," accessed Oct. 1, 2015.

[6] I. Ahmad, X. Wei, Y. Sun, Y.-w.Q. Zhang, "Video transcoding: an overview of various techniques and research issues", *IEEE Transactions on Multimedia*, vol. 7, no. 5, pp. 793–804, 2005.

[7] G. Gao, and Y. Wen, "Video transcoding for adaptive bitrate streaming over edge-cloud continuum," *Digital Communications and Networks*, vol. 7, no. 4, pp. 598-604, 2021.

[8] X. Li, M.A. Salehi, Y. Joshi, M.K. Darwich, B. Landreneau, and M. Bayoumi, "Performance analysis and modeling of video transcoding using heterogeneous cloud services," *IEEE Transactions on Parallel and Distributed Systems*, vol. 30, no. 4, pp. 910-922, 2018.

[9] S.M. Farhad, M.S.I. Bappi, and A. Ghosh, "Dynamic resource provisioning for video transcoding in iaas cloud," *In 2016 IEEE 18th International Conference on High Performance Computing and Communications; IEEE 14th International Conference on Smart City; IEEE 2nd International Conference on Data Science and Systems (HPCC/SmartCity/DSS)*, pp. 380-384, 2016.

[10] S.M.A.H. Bukhari, K. Bilal, A. Erbad, A. Mohamed, and M. Guizani, "Video transcoding at the edge: cost and feasibility perspective," *Cluster Computing*, pp.1-24, 2022.

[11] Opinion Model for Network Planning of Video and Audio Streaming Applications, document *ITU-T Recommendation* G.1071, Jun. 2015.

[12] Parametric Bitstream-Based Quality Assessment of Progressive Download and Adaptive Audiovisual Streaming Services Over Reliable Transport-Video Quality Estimation Module, document *ITU-T Recommendation P.1203.1*, Oct. 2017.

[13] Video Quality Experts Group (VQEG). [Online]. Available: https://www.its.bldrdoc.gov/vqeg/projects/audiovisual-hd.aspx.

[14] Y. Wang, S. Inguva, and B. Adsumilli, "YouTube ugc dataset for video compression research," *in IEEE International Workshop on Multimedia Signal Processing*,2019.

[15] A. Vetro, C. Christopoulos, and H. Sun, "Video transcoding architectures and techniques: an overview," *IEEE Magazine on Signal Processing*, vol. 20, no. 2, pp. 18–29, Mar. 2003.

[16] X. Li, M.A. Salehi, Y. Joshi, M.K. Darwich, B. Landreneau, and M. Bayoumi, "Performance analysis and modeling of video transcoding using heterogeneous cloud services," *IEEE Transactions on Parallel and Distributed Systems*, 30(4), pp.910-922, 2018.

[17] T. Wiegand, G. J. Sullivan, G. Bjontegaard, and A. Luthra, "Overview of the h. 264/avc video coding standard," *IEEE Transactions on circuits and systems for video technology*, vol. 13, no. 7, pp. 560–576, 2003.

[18] F. Lao, X. Zhang, and Z. Guo, "Parallelizing video transcoding using map-reduce-based cloud computing," *in Proceedings of IEEE International Symposium on Circuits and Systems*, ser. ISCAS '12, pp. 2905–2908, 2012.

[19] H. Sun, W. Kwok, and J. Zdepski, "Architectures for MPEG compressed bitstream scaling," *IEEE Transactions on Circuits and Systems for ideo Technology*, vol. 6, no. 2, pp. 191–199,1996

[20] T.-K. Lee, C.-H Fu, Y.-L. Chan, and W.-C. Siu, "A new motion vector composition algorithm for fast-forward video playback in H.264," *in IEEE International Symposium on Circuits and Systems*, pp. 3649–3652), 2010.

[21] T. Shanableh and M. Ghanbari, "Heterogeneous video transcoding to lower spatiotemporal resolutions and different encoding formats," *IEEE Transactions on Multimedia*, vol. 2, pp. 101–110, 2000

[22] J. Xin, C.-W. Lin, and M.-T. Sun, "Digital video transcoding," *Proceedings of the IEEE*, vol. 93, no. 1, pp. 84–97, 2005.

[23] P. Kunzelmann and H. Kalva, "Reduced complexity H.264 to MPEG- 2 transcoder," *IEEE International Conference on Consumer Electronics (ICCE 2007)*, pp. 12, 2007.

[24] H. Shu and L.-P. Chau, "The realization of arbitrary downsizing video transcoding," *IEEE Transactions on Circuits and Systems for Video Technology*, vol. 16, no. 4, pp. 540–546, 2006.

[25] D. Zhang, B. Li, J. Xu, and H. Li, "Fast Transcoding from H. 264 AVC to High Efficiency Video Coding," *IEEE International Conference on Multimedia and Expo (ICME)*, pp. 651-656, 2012.

[26] D. Seo, J. Kim, and I. Jung, "Load distribution algorithm based on transcoding time estimation for distributed transcoding servers," *in Proceedings of International Conference on Information Science and Applications*, ser. ICISA '10, pp. 1–8, 2010.

[27] A.S. Nagaraghatta, "Algorithms and methods for video transcoding (Doctoral dissertation)", 2019.

[28] L.P. Van, J. De Praeter, G.Van Wallendael, S. Van Leuven, J. De Cock, and R. Van de Walle, "Efficient bit rate transcoding for high efficiency video coding," *IEEE Transactions on Multimedia*, 18(3), pp.364-378, 2015.

[29] O. Werner, "Requantization for transcoding of MPEG-2 intraframes," *IEEE Transactions on Image Processing*, vol. 8, pp. 179–191, 1999.

[30] J. Jiang, V. Sekar, and H. Zhang, "Improving fairness, efficiency, and stability in http-based adaptive video streaming with festive," *in Proceedings of the 8th International Conference on emerging Networking Experiments and Technologies*, pp. 97–108, 2012.

[31] P. Assuncao and M. Ghanbari, "A frequency-domain video transcoder for dynamic bit-rate reduction of MPEG-2 bit streams," *IEEE Trans. Circuits Syst. Video Technolgy.*, vol. 8, no. 8, pp. 953–967, 1998.

[32] A. Eleftheriadis and P. Batra, "Dynamic rate shaping of compressed digital video," *IEEE Trans. Multimedia*, vol. 8, no. 2, pp. 297–314, 2006.

[33] J. De Cock, S. Notebaert, P. Lambert, and R. Van de Walle, "Requantization transcoding for H.264/AVC video coding," *Signal Processing: Image Communication*, vol. 25, pp. 235–254, 2010.

[34] N. Hait and D. Malah, "Model-based transrating of H.264 coded video," *IEEE Trans. Circuits Syst. Video Technol.*, vol. 19, no. 8, pp. 1129–1142, 2009.

[35] N. Bjork and C. Christopoulos, "Transcoder architectures for video coding," *IEEE Transactions on Consumer Electronics*, vol. 44, no. 1, pp. 88–98, 1998.

[36] D. Shen, I. K. Sethi, and B. Vasudev, "Adaptive motion-vector resampling for compressed video downscaling," *IEEE Trans. Circuits and Systems for Video Technology*, vol. 9, no. 6, p.929, 1999.

[37] T. Shanableh and M. Ghanbari, "Heterogeneous video transcoding to lower spatio-temporal resolutions and different encoding formats," *Multimedia, IEEE Transactions on*, vol. 2, no. 2, pp. 101 -110, 2000.

[38] P. Yin, A. Vetro, B. Liu, and H. Sun, " Drift compensation for reduced spatial resolution transcoding," *IEEE Transactions on Circuits and Systems for Video Technology*, vol. 12, no. II, pp. 1009 - 1020, 2002.

[39] Y. Sambe, S. Watanabe, D. Yu, T. Nakamura, and N. Wakamiya, "High-speed distributed video transcoding for multiple mtes and formats," *IEICE Transactions*, vol. 88-D, no. 8, pp. 1923-1931, 2005.

[40] B Hu, P Zhang, Q Huang, and W Gao, "Reducing Spatial Resolution for MPEG-2 to H.264/AVC Transcoding," *In Advances in Multimedia Information Processing-PCM 2005: 6th Pacific Rim Conference on Multimedia,* pp. 830-840,2005

[41] Sung-Eun Kim, Jong-Ki Han, and Jae-Gon Kim, "Efficient Motion Estimation Algorithm for MPEG-4 to H.264 Transcoder," *IEEE International Conference on Image Processing,* vol. 3, pp. III-656, 2005

[42] S. Goel, Y. Ismail, and M. Bayoumi, "High-speed motion estimation architecture for real-time video transmission," *The Computer Journal*, vol. 55, no. 1, pp. 35–46, 2012.

[43] M.A. Bonuccellit, F. Lonetti, and F. Martelli, "Temporal transcoding for mobile video communication," *In the second Annual International Conference on Mobile and Ubiquitous Systems: Networking and Services,* pp. 502-506. 2005

[44] B. G. Haskell, A. Puri, and A. N. Netravali, Digital video: an introduction to MPEG-2. Springer Science and Business Media, 1996.

[45] G. J. Sullivan, J.-R. Ohm, W.-J. Han, and T. Wiegand, "Overview of the high efficiency video coding (hevc) standard," *IEEE Transactions on circuits and systems for video technology*, vol. 22, no. 12, pp. 1649–1668, 2012.

[46] M. Shaaban and M. Bayoumi, "A low complexity inter mode decision for MPEG-2 to H.264/avc video transcoding in mobile environments," *in Proceedings of the 11th IEEE International Symposium on Multimedia*, pp. 385–391, 2009.

[47] A. Ortega, and K. Ramchandran, "Rate-distortion methods for image and video compression," *IEEE Signal processing magazine*, 15(6), pp.23-50, 1998.

[48] J. MacQueen, "Some methods for classification and analysis of multivariate observations," *In Proceedings of the fifth Berkeley symposium on mathematical statistics and probability,* Vol. 1, No. 14, pp. 281-297, 1967.

[49] L. Prangnell, "Visually lossless coding in HEVC: A high bit depth and 4: 4: 4 capable JND-based perceptual quantization technique for HEVC," *Signal Processing: Image Communication*, 63, pp.125-140, 2018.